\documentclass[twocolumn]{svjour3}          
\pdfoutput=1
\usepackage{natbib}
\usepackage{graphicx}
\usepackage{subfigure}
\usepackage[usenames,dvipsnames,svgnames,table]{xcolor}
\usepackage[show]{ed}

\usepackage{hyperref}
\usepackage[english]{babel}

\usepackage[utf8]{inputenc}
\usepackage[T1]{fontenc}
\usepackage{textcomp}
\usepackage{lmodern}

\usepackage{listings}

\begin{document}

\title{Building ArtBots to Attract Students into STEM Learning}
\subtitle{A Case Study of an International Robot Week for Secondary School Children}


\author{Francis wyffels \and Willem Van de Steene \and Jelle Roets \and Maria-Cristina Ciocci \and Juan Pablo Carbajal
}


\institute{Francis wyffels \at
              Ghent University, Dept. of Electronics and Information Systems, and Dwengo \\
              \email{Francis.wyffels@UGent.be}           
           \and
           Willem Van de Steene \at
              Ghent University, Dept. of Materials Science and Engineering
            \and
            Jelle Roets \at
               Autodesk and Dwengo
            \and 
            Maria-Cristina Ciocci \at 
               Ghent University, Dept. of Information Technology, and Ingegno 
               \and 
            Juan Pablo Carbajal \at 
               Dwengo Helvetica
}



\maketitle

\begin{abstract}
There is an increasing worldwide demand for people educated into science and technology. Unfortunately, girls and underprivileged students are often underrepresented in Science, Technology, Engineering and Mathematics (STEM) education programs. 
We believe that by inclusion of art in these programs, educational activities might become more attractive to a broader audience. 
In this work we present an example of such an educational activity: an international robotics and art week for secondary school students.
This educational activity builds up on the project-based and inquiry learning framework. This article is intended as a brief manual to help others organise such an activity.
It also gives insights in how we led a highly heterogeneous group of students into  learning STEM and becoming science and technology ambassadors for their peers.
\end{abstract}

\section{Introduction}

The rise of the Makers Movement is one of the most exciting trends in the past decade~\citep{Martin2015}.
In a learning environment, a maker is a true learner that values the process of making as much as the product.
Making has the potential to create life-long learners and it is an activity that fits perfectly within Project Based Learning (PBL) curricula~\citep{bender2012PBL}.

In this context, hands-on robotics appears as an ideal instructional approach~\citep{Rieber1996,Catlin2012}.
The impact of a hands-on robotics program focused on expression and creativity in the service of improving student technological fluency has been recently highlighted~\citep{Hamner_2013_7442}.
Firstly, such a program offers (classroom) activities that expose students to valuable concepts from Science, Technology, Engineering and Mathematics (STEM), and it allows to address STEM topics deemed strategic in several national education programs.
Secondly, these activities are aligned with core constructionism concepts (e.g. ``think with your hands'') and with current neuroscience research on teaching to the brain’s natural learning system~\citep{given2002}.
Finally, activities involving robotics also foster essential skills like problem solving, collaboration and project management.
The last two skills are specially relevant when participants work in groups, as done in scientific and engineering work.

The natural curiosity in kids and teenagers is particularly strong when it comes to machines.
This make robots ideal tools to keep them engaged and motivated during learning.
This is specially true when the learning methodology uses unconstrained robotics, i.e. building robots without preassembled kits.
Unconstrained robotics truly puts the participant in the role of the maker, bringing them to decision-making situations that accurately simulate working environments. 
For example, a group must first agree on a robot design; this implies the exchange of explanations to illustrate each member's ideas and decide on the contributions that will be part of the final design.
Once a design is in place, the group needs to effectively distribute tasks among its members and make time scheme.

Robotics activities are well scaffolded with inquiry project-based learning, which sets up an environment where participants are engaged in open-ended, student-centered, hands-on activities~\citep{Colburn2000}.
This involves tackling real-world problems and generating solutions, which can also be relevant for the participants' community~\citep{Bouillion2001}.
In this environment, the learning process is shared with the tutors who guide and stimulate participants to question~\citep{freire1989question}.

Despite the fact that building a robot is an exiting idea (probably due to its strong presence in popular culture), it might not resonate with the interest of all students.
Noteworthy, there are gender issues to be addressed in robotics or more generically in STEM.
Worldwide girls are underrepresented in STEM oriented higher education programs and professions~\citep{burke2007}. Among other things, this segregation is troublesome for industries: when women are not involved in the design of their products, needs and desires unique to women may be overlooked~\citep{Hill2010}\footnote{The case of LEGO is analysed in \url{https://youtu.be/CrmRxGLn0Bk}}.
The gender difference is dramatic in well developed countries~\citep{Sjoberg2010} and there seems to be a relation with how technology is perceived by girls, which discourages them to pursue activities in the topic.
Evidence shows that robots equally attract girls and boys, but each gender might be captivated by different aspects of the technology.
A solution is to design robotic activities that are attractive to both genders~\citep{Johnson2003} and this might be catalysed by the inclusion of art.

In this paper we discuss a four days long robotics and art activity (ArtBots) organized for a highly heterogeneous group of teenagers.
During ArtBots, students were asked to design and build their own robot artist.
The activity involved the interaction between participants of different ages, socio-economical backgrounds and cultures.
The physical results of the process, the ArtBots, were exposed on the last day during an exhibition at a Belgian contemporary art museum.
The goals of the activity were threefold:

\begin{enumerate}
    \item To provide technical skills and robot building experience to high-school students,
    \item To cause a cross-cultural exchange of knowledge and experiences,
    \item To train technology ambassadors who stimulate friends and other students to get involved in technology.
\end{enumerate}

In this article, see Section~\ref{sec:introbotweek}, we document the activity so that it could be reproduced by others. Additionally, we discuss ArtBots as an inquiry project-based learning approach for STEM and we share our insights on how the project affected the way participants perceive STEM (see Section~\ref{sec:discussion}).

\section{An international robot week for secondary school students}\label{sec:introbotweek}

\begin{figure}[t]
    \centering
    \includegraphics[width=0.45\textwidth]{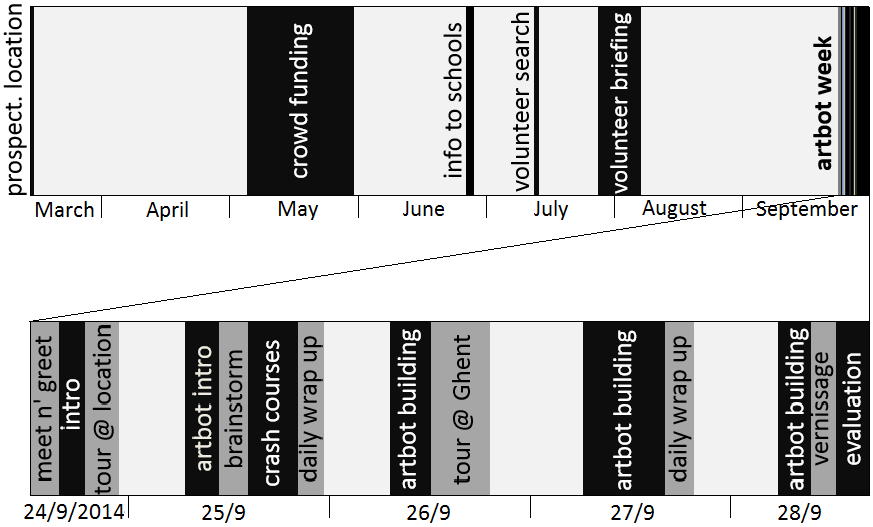}
    \caption{Timeline of the organization and execution of ArtBots. The event lasted 4 days while its preparation required the orchestrated actions of volunteers during 7 months: $98\%$ of the time was spent in the preparation of the event.}\label{fig:timeline}
\end{figure}

While the ArtBots event itself spanned only 4 days (from Thursday evening until Monday afternoon), preparation took much longer.
It involved a crowdfunding campaign, searching a workspace and sleeping accommodation, acquiring material and tools, informing the participants, their schools and their families, and much more.
An overview of the time distribution can be found in Figure~\ref{fig:timeline}.

In the following sections we focus on the selection of the participants and the execution of the ArtBots event itself.

\subsection{Selecting the participants}
ArtBots was intended for teenagers (from 12 to 18 years old) with a socio-economically disadvantaged background and a strong interest in science and technology.
Additionally, we aimed for gender equality.
In order to reach out to this audience, we defined a set of guidelines for the selection of students that were sent to schools and other organisations that we contacted:

\begin{itemize}
    \item Age: students were selected in the age range of 12 to 18 year (high school).
    \item Gender: we aimed for a healthy collaboration between girls and boys.
    We suggested schools and organisations to assemble gender-balanced groups and to give preference to lesbian, gay, bisexual, and transgender (LGBT) students.
    \item Motivation: we asked for students who were fascinated by science and technology, eager to design and build machines or to program applications.
    \item Social background: different social backgrounds were described.
    The preference was given to students with no opportunities to travel nor access to other leisure activities (children's camps) during their holidays.
    Students from economically disadvantaged households should be treated with preference.
    \item Language: participants had to have basic English understanding (level A1 of the Common European Framework of Reference for Languages~\citep{Europe2004}).
\end{itemize}

\begin{figure}[t]
    \centering
    \includegraphics[width=0.45\textwidth]{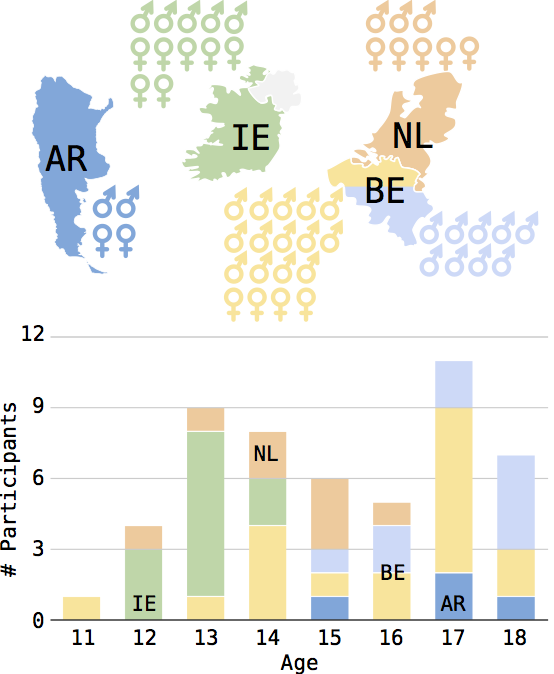}
    \caption{Summary of participants' count (51 in total), gender and nationality. Credits: Maps from Wikimedia}\label{fig:participants}
\end{figure}

The above criteria served as guidelines only.
Figure~\ref{fig:participants} summarizes the characteristics of the participants.
Most of them had little experience with both art and robots.
We did not survey the actual participants for their social backgrounds. Consequently, we can not give any quantitative results on this matter.
However, we observed that we had a good mixture (approx. $50\%$) of underprivileged and privileged teenagers.

\subsection{Art and robots\label{ArtBots}}

\begin{figure}[t]
    \centering
    \begin{subfigure}
        \centering
        \includegraphics[height=1.5in]{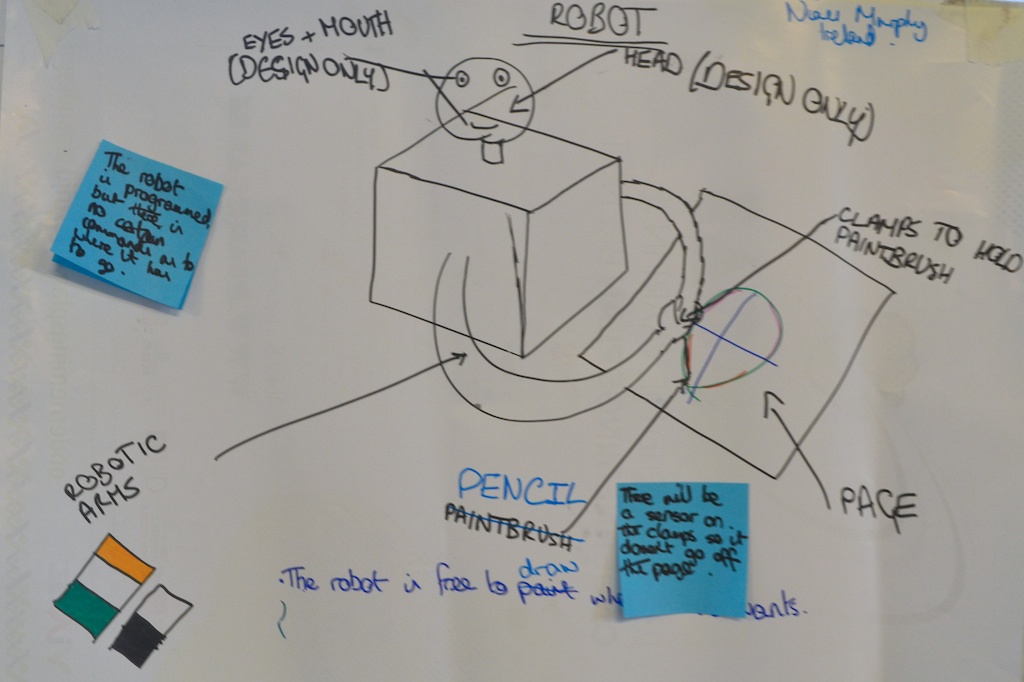}
    \end{subfigure}
    \begin{subfigure}
        \centering
        \includegraphics[height=1.5in]{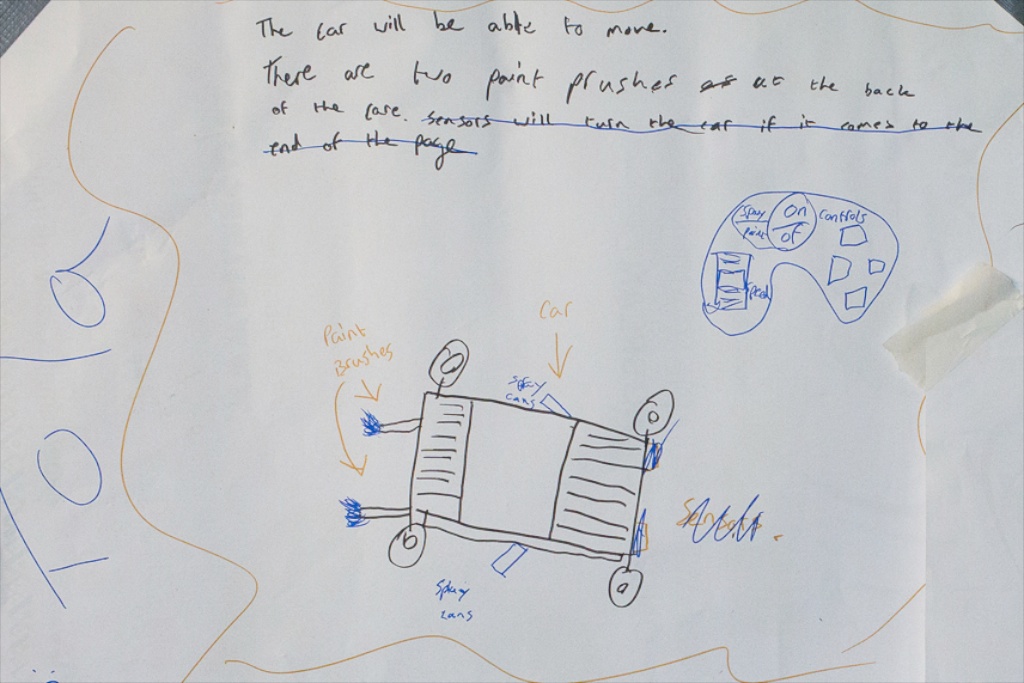}
    \end{subfigure}
    \begin{subfigure}
        \centering
        \includegraphics[height=1.5in]{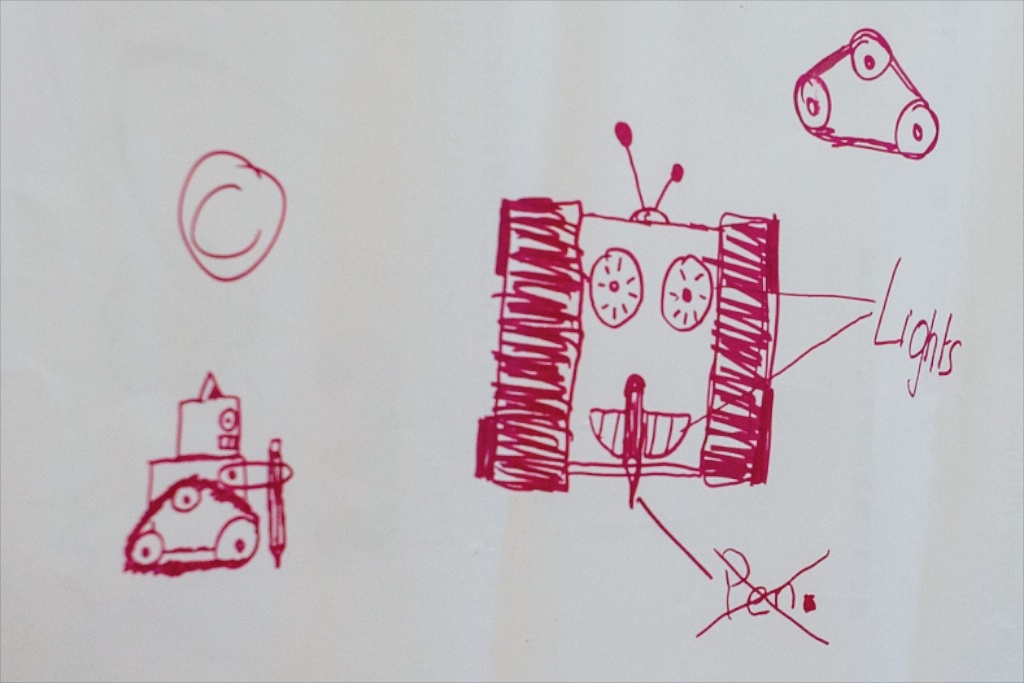}
    \end{subfigure}
    \caption{Some of the participants' robot designs after the brainstorming session with the artists.}\label{fig:robot_sketches}
\end{figure}

The venue took place at a private art site 
with a large modern and contemporary art exhibition space.
The inspiring and primitive environment immediately put our participants in an artistic mindset.
In order to give the participants a better understanding about art, we collaborated with an artist co-creation platform 
who sent two artists to present their views on art.
The artists organised a brainstorm session where participants had to express their views and ideas about artistic robots.
The key idea was to give each participant the time to think about their own ArtBot without being constrained by tools, materials or knowledge.
During this brainstorm process, all participants expressed their ideas on paper. This is illustrated in Figure~\ref{fig:robot_sketches}.

\subsection{Crash course into programming, electronics and mechanics}
Robotics emerges from the combination of electronics, mechanics and programming.
To build a robot, the designer must get in touch with all these fields.
During ArtBots we provided crash courses to all the participants.
Each crash course was designed and taught by volunteers with experience in the field.

\subsubsection{Programming and algorithmic thinking\label{program}}
It is unrealistic to think that we could teach programming in a few hours. However, the materials used for this event, namely the Dwenguino board\footnote{The Dwenguino board is a multi-functional Arduino IDE compatible microcontroller platform (\url{http://www.dwengo.org/tutorials/dwenguino/dwenguino-board}).}, 
allowed us to focus on a reduced set of programming and algorithmic skills: conditionals, loops and functions.
This reduced skill set was sufficient to program robots with complex behaviors.

To introduce the participants to conditionals and loops we used the Blockly Game suit\footnote{https://blockly-games.appspot.com}.
After a short and playful introduction to the topic and a short icebreaker Q\&A session, we invited the participants to set the interface to the language they felt most comfortable with and then we asked them to solve the ``Maze'' challenges of Google Blockly.
In these challenges the user must write instructions for a software agent, such that it moves through a maze from an initial point to a final destination marked with an arrow.
Participants with programming experience were invited to opt for harder problems from Google Blockly.
The most experienced participants settled into the programming games ``Turtle'' and ``Pond''.
We let them play with the game for about 45 minutes.

The second part of the crash course focused on the use of the Dwenguino board and took approximately 30 minutes.
In this part we also introduced the idea of using functions.
We used the integrated LCD screen to show how functions are used.
We allowed free exploration of the LCD and the other functions of the Dwenguino.
At the same time, small challenges were provided to guide students that might felt lost.
For example, we asked them to write code to turn on the backlight of the LCD, to blink a LED, to detect a pressed button, etc.
A small sample source code of an easy challenge would look like the one shown in Snippet~\ref{code:example}. 

\lstset{language=C++,
                basicstyle=\ttfamily,
                keywordstyle=\color{blue}\ttfamily,
                stringstyle=\color{red}\ttfamily,
                commentstyle=\color{gray}\ttfamily,
                morecomment=[l][\color{magenta}]{\#}
}
\begin{lstlisting}[caption={Code used for the LCD programming challenge.},label=code:example]
#include <LiquidCrystal.h>
#include <Wire.h>
#include <Dwenguino.h>

void setup() {
 initDwenguino();
 // Use this function to print your name
 dwenguinoLCD.print("Hello!");
 // Can you read it?
 // How do you turn on the backlight?
}

void loop() {}
\end{lstlisting}

Finally, we offered the participants a rather complex source code sample that printed the letter "A" on the LCD and blinked a LED every time a button was pressed.
We asked them to discover what part of the code was responsible for the blinking LED and we challenged them to modify the code such that the letter "A" moved one place to the left each time the button was pressed.
This {\it find and hack} task was optional and meant to encourage deeper understanding of the code.

\subsubsection{Electronics\label{electron}}
Similar to the programming course, the electronics crash course was given in a hands-on way.
The goal of this course was to provide the basic concepts of electronics with respect to robotics rather than overwhelming the participants with theory.

Initially, participants had to figure out what is the use of electronics.
For example, if they wanted to make a painting robot, we ask them to think about how would they tackle the challenge by themselves.
This guided them to analyse the painting behavior.
Afterwards, they had to translate this into a robot and its components.
In order words, they had to think about electronic components as equivalents of human senses, brain and muscles.
The participants were able to come up with examples such as sound sensors, light sensors, push buttons and ultrasonic sensors.
When discussing about the robot's intelligence, participants proposed the analogy between the brain and a programmable computer (or microcontroller).
Servomotors and other type of motors would play the role of muscles.


After this initiation, the main part of the crash course took place which consisted of solving different challenges in small groups of two or three participants.
For example, one of these challenges consisted in controlling the position of a servo with a potentiometer.
This is a relatively simple task that covers all the necessary concepts: analog input processed by a microcontroller which controls actuators.
We provided all the components (sensors, the Dwenguino board and motors) and an incomplete source code.
Participants had to build the electronic circuit and complete the code.
Additional information and answers to questions were provided on demand, by the tutoring volunteers.

\subsubsection{Mechanics}
For the mechanical part of the robot development, we supplied participants with a broad range of materials and tools.
We set up a live demonstration and try-out for both hand and power tools available in the workshop.
To learn the names of the tools available in the workshop, overcoming the language diversity of the group, we organized games.
We did this to avoid giving an extensive enumeration and description of each tool.
We also discussed safety measures for correct handling of the tools.

Following up to this introduction, we elaborated on appropriate material and tool selection for different applications, as well available materials for joining techniques such as (hot) gluing, bolting, screwing, nailing and interlocking.

When prototyping a robot, Computer Aided Design (CAD) can be used for a better 3D visualisation and communication of different concepts and ideas.
To introduce the basics of 3D modeling, we gave a short demonstration followed by individual practice in redrawing simple existing objects.
We used Autodesk's 123D Design\footnote{A basic and intuitive freeware CAD tool \url{http://www.123dapp.com/design}} for designing some of the more complex robots' parts. 
The majority of those custom developed parts were later produced through additive manufacturing, also known as 3D printing.
To warrant success of the 3D printing process, participants received easy-to-understand guidelines.

\subsection{From idea to realisation}

\begin{figure}[t]
    \centering
    \begin{subfigure}
        \centering
        \includegraphics[height=1.5in]{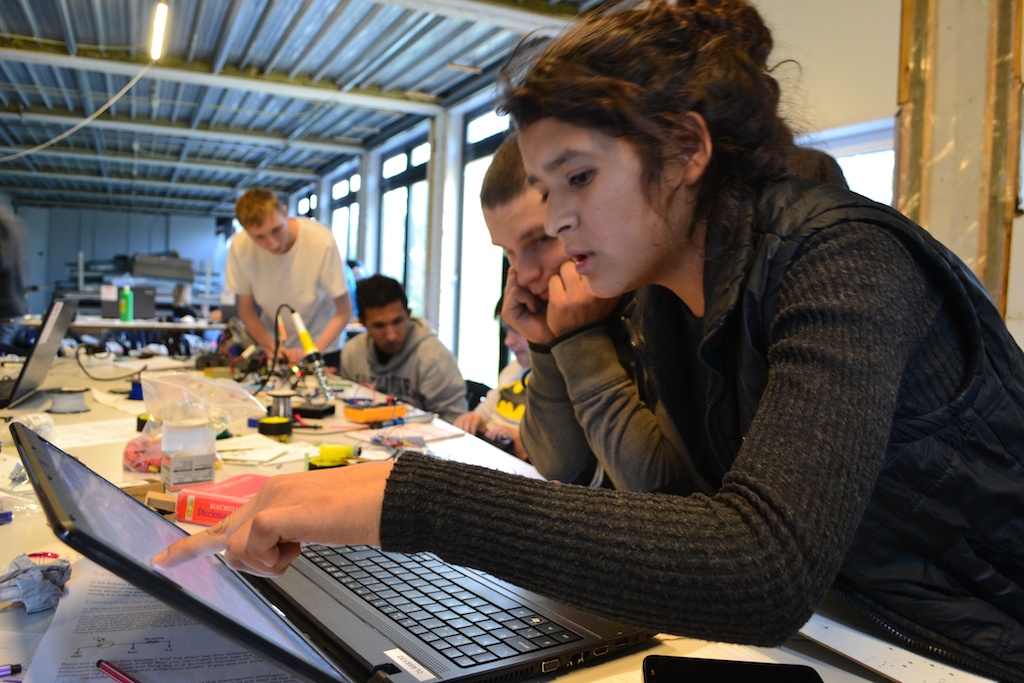}
    \end{subfigure}
    \begin{subfigure}
        \centering
        \includegraphics[height=1.5in]{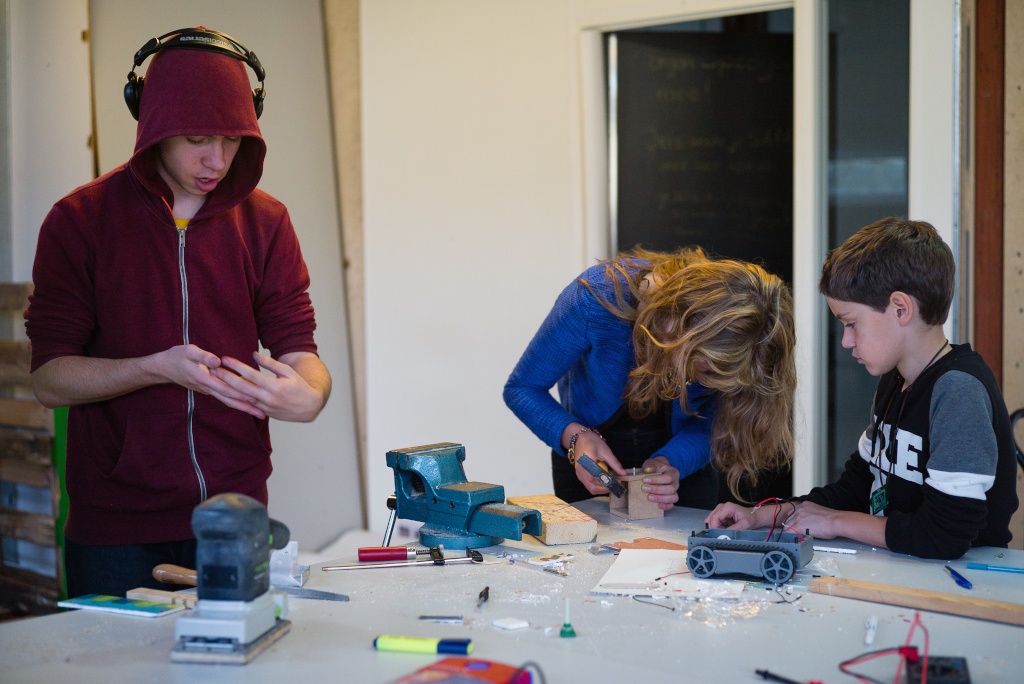}
    \end{subfigure}
    \begin{subfigure}
        \centering
        \includegraphics[height=1.5in]{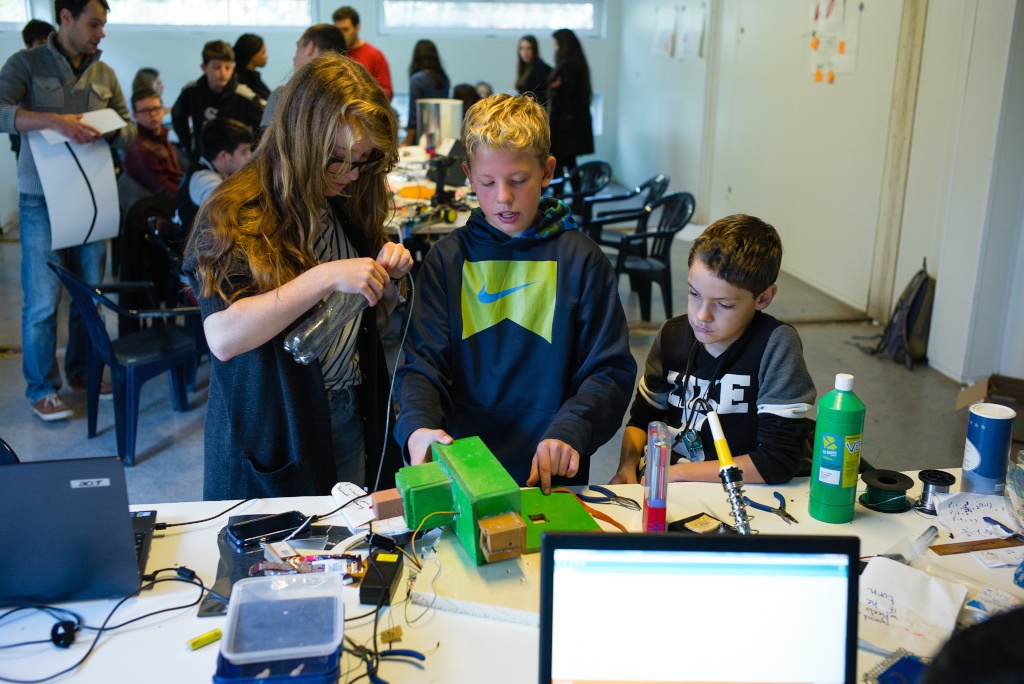}
    \end{subfigure}
    \caption{Some photographs taken during the building process}\label{fig:robot_building}
\end{figure}

\begin{figure}[ht]
    \centering
    \begin{subfigure}
        \centering
        \includegraphics[height=1.5in]{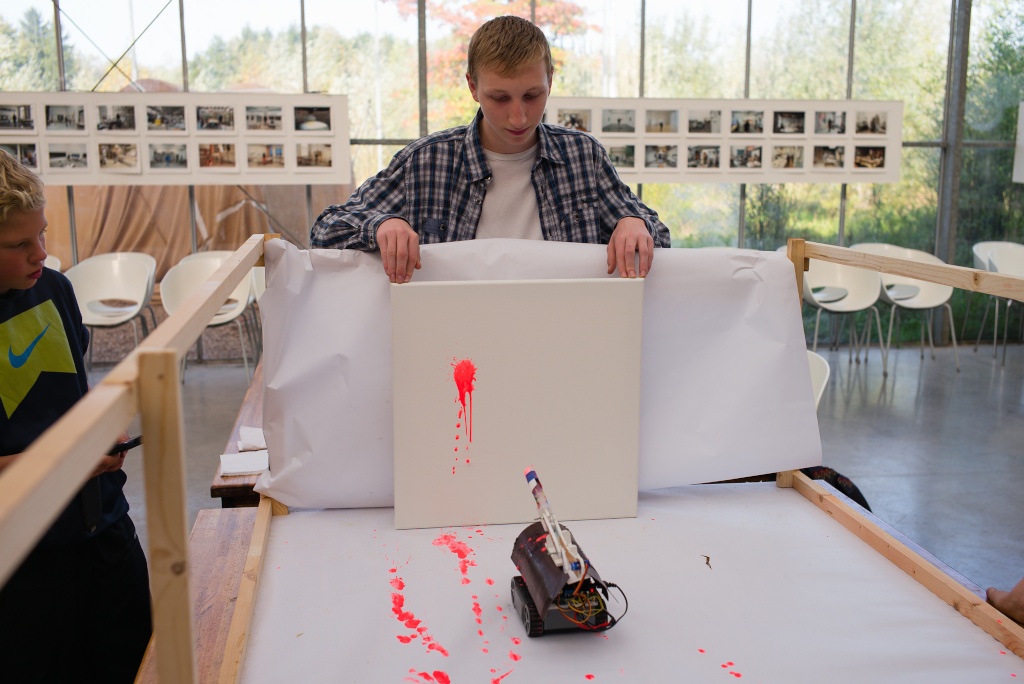}
    \end{subfigure}
    \begin{subfigure}
        \centering
        \includegraphics[height=1.5in]{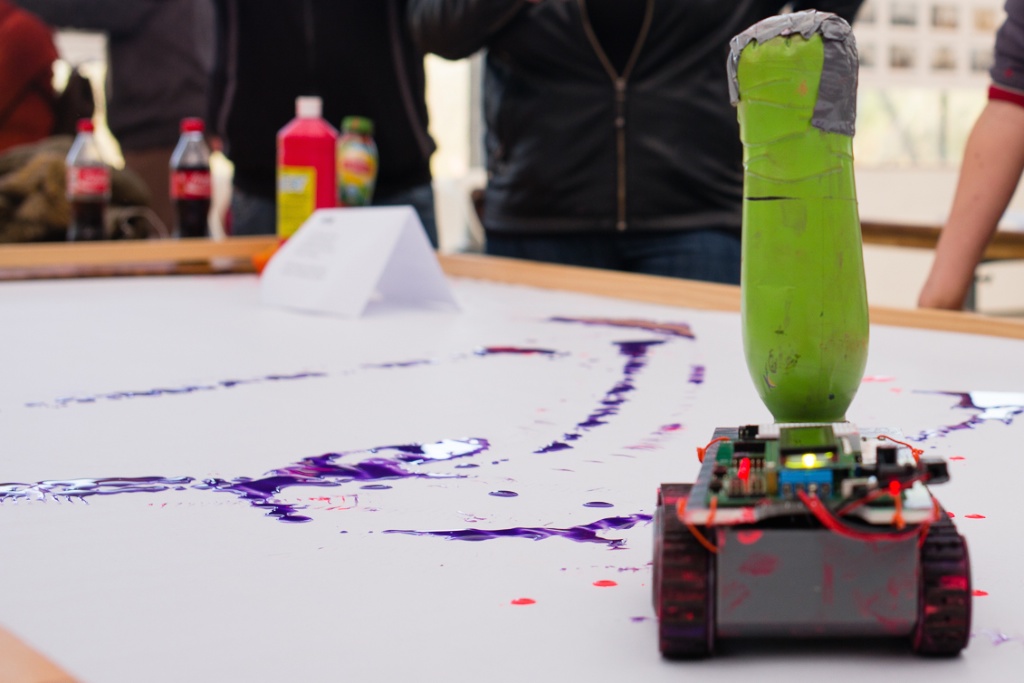}
    \end{subfigure}
    \begin{subfigure}
        \centering
        \includegraphics[height=1.5in]{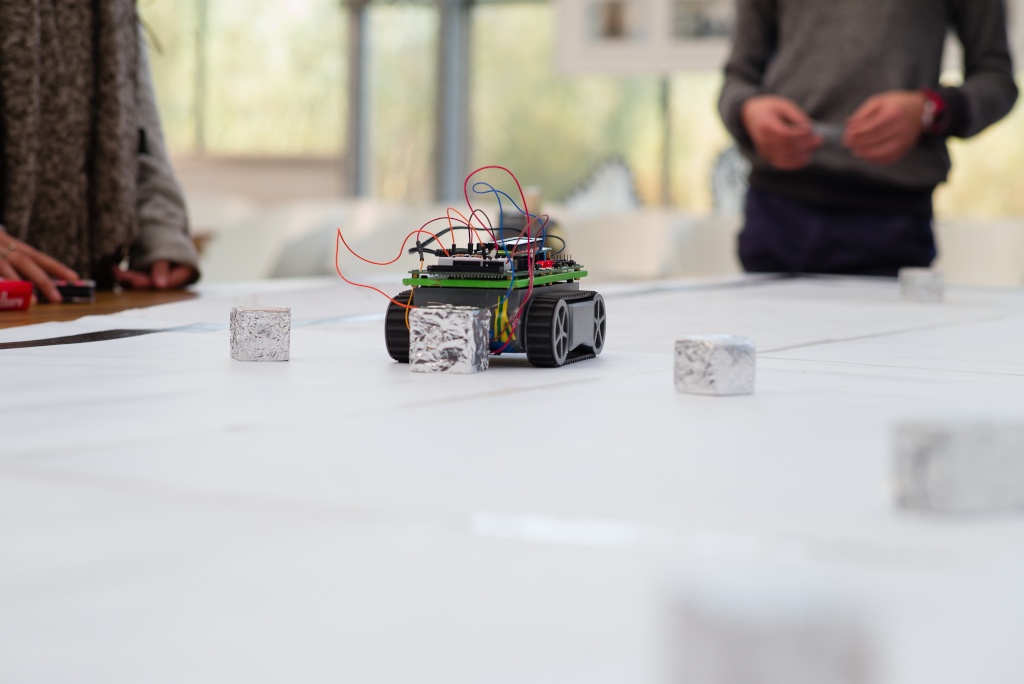}
    \end{subfigure}
    \begin{subfigure}
        \centering
        \includegraphics[height=1.5in]{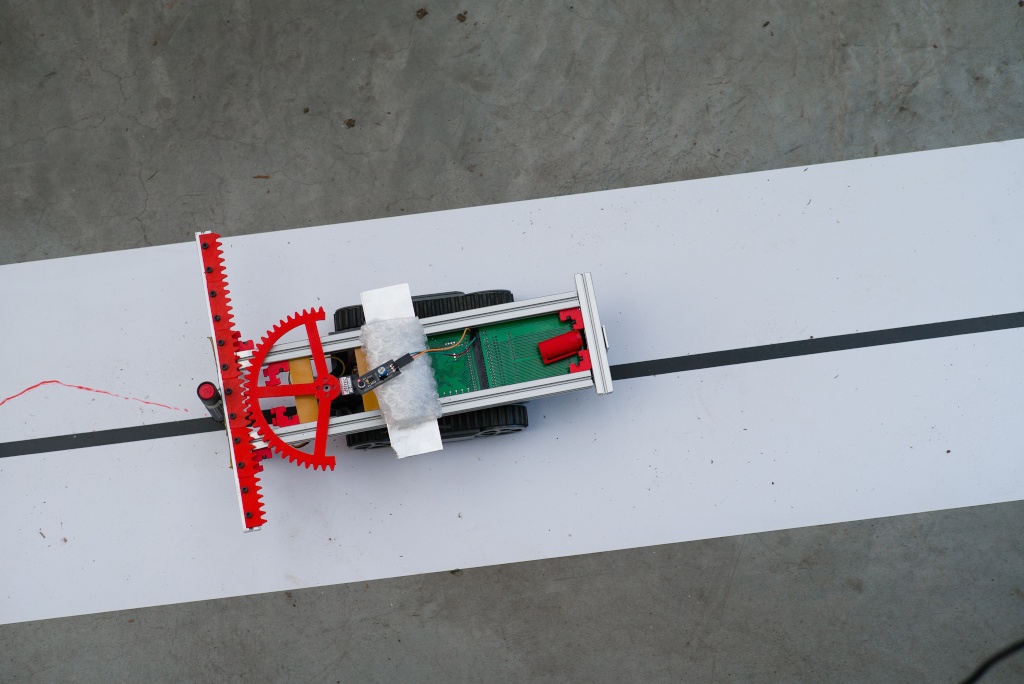}
    \end{subfigure}
    \caption{A selection from the realised ArtBots: a paint shooting robot, a dripping robot, a block collecting robot and a sound registering robot.}\label{fig:robot_selections}
\end{figure}

During the crash courses, we, the organisers, gathered and clustered the robot designs (i.e. the sketches) based on the concepts within the design.
For example, multiple participants wanted to build a robot that integrated some action painting (e.g.~throwing or shooting paint).
And thus, we put together all ideas related to action painting.
The clustering process was performed without any interference from participants.
Consequently, the clustering did not consider age, cultural background or gender of the participants.
This resulted in 12 clusters and 12 teams, with ideas varying from action painting to abstract interactive ArtBots.

After the crash courses, we asked the participants to find their team mates within the assigned clusters.
Next, we asked them to refine their ideas about programming, electronics and mechanics with respect to their robot designs, and start the building process.
Mechanical components, motors, microcontroller boards and sensors were provided to the participants as basic building blocks for their robot.
Additionally, participants had access to all the necessary tools to build their robot.
This included computers, 3D printers, milling machines, soldering stations etc.
As the timeline in Figure~\ref{fig:timeline} shows, the building process was spread over three days and took the largest part of the available time frame.
Action pictures of the students during the building process are given in the pictures on Figure~\ref{fig:robot_building}.

Although some groups had to work until the very last minute, all 12 teams managed to finishing their robot.
Most robots could paint by using stamps, brushes, dripping and shooting mechanisms, while some other robots were more abstract, including interactive poetic robots and a block collecting robot. Figure~\ref{fig:robot_selections} gives four examples.


\subsection{Social activities}
In order to provide some distraction, social activities were organised which varied from short in-between activities such as taking a 3D scan of your face and 3D printing it, and playing sports games. Additionally, we organised some longer activities such as an afternoon visit to the historical town centre of Ghent.

\subsection{Exhibition}
\begin{figure*}[t]
    \centering
    \begin{subfigure}
        \centering
        \includegraphics[height=1.5in]{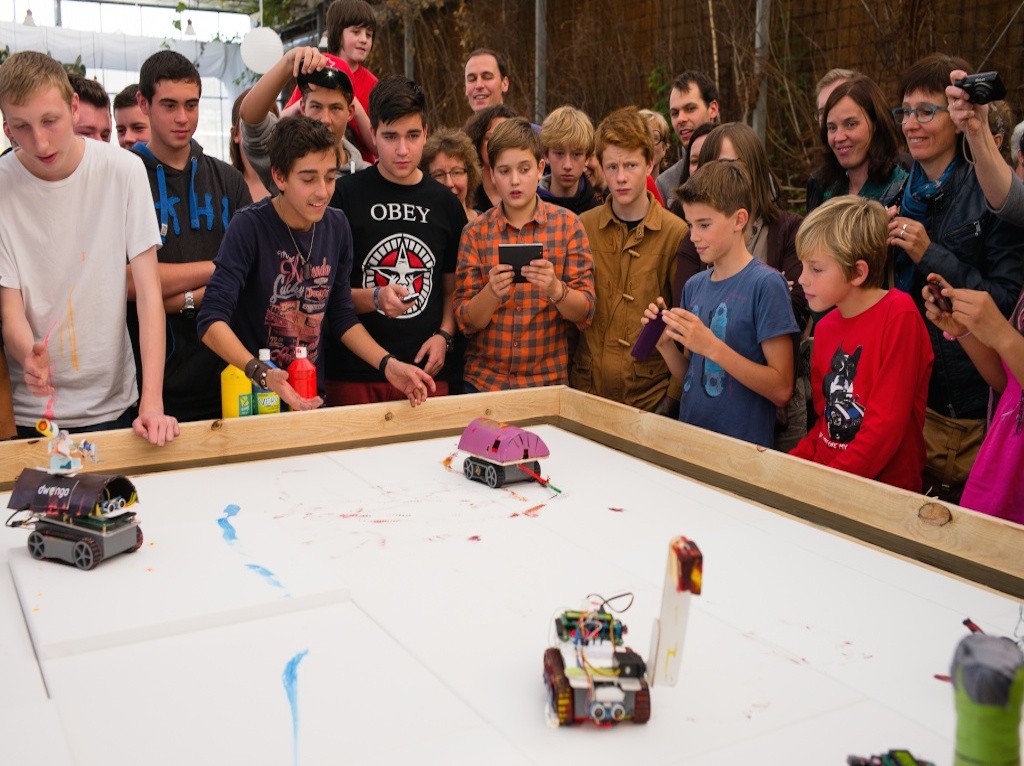}
    \end{subfigure}
    \begin{subfigure}
        \centering
        \includegraphics[height=1.5in]{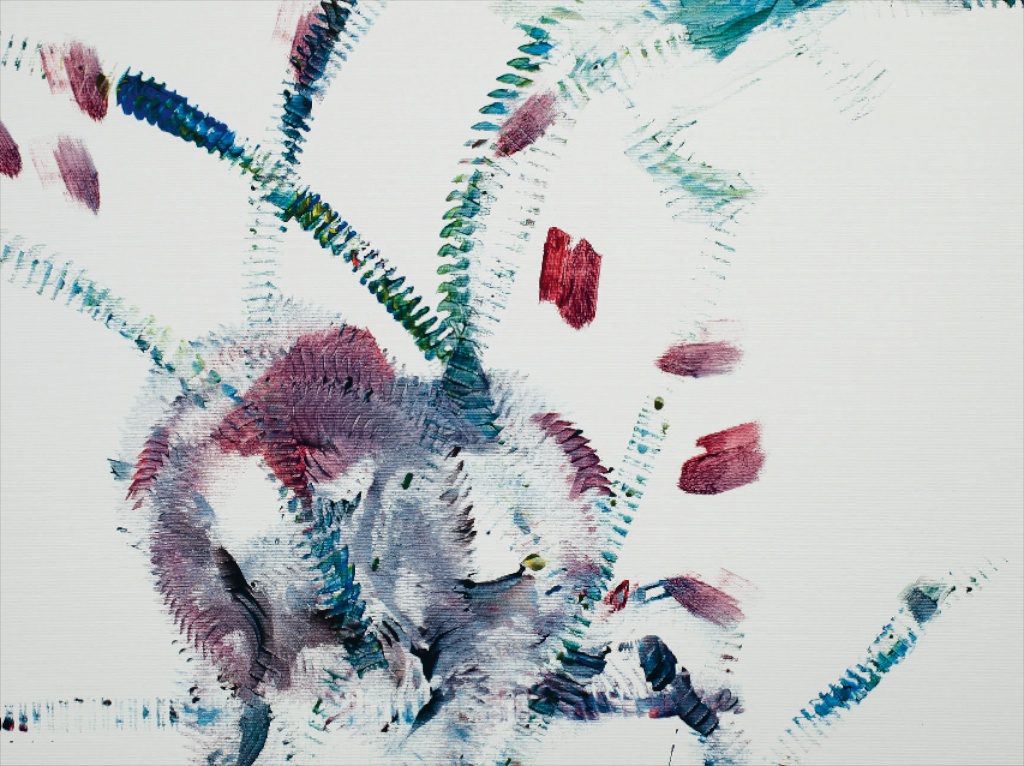}
    \end{subfigure}
    \begin{subfigure}
        \centering
        \includegraphics[height=1.5in]{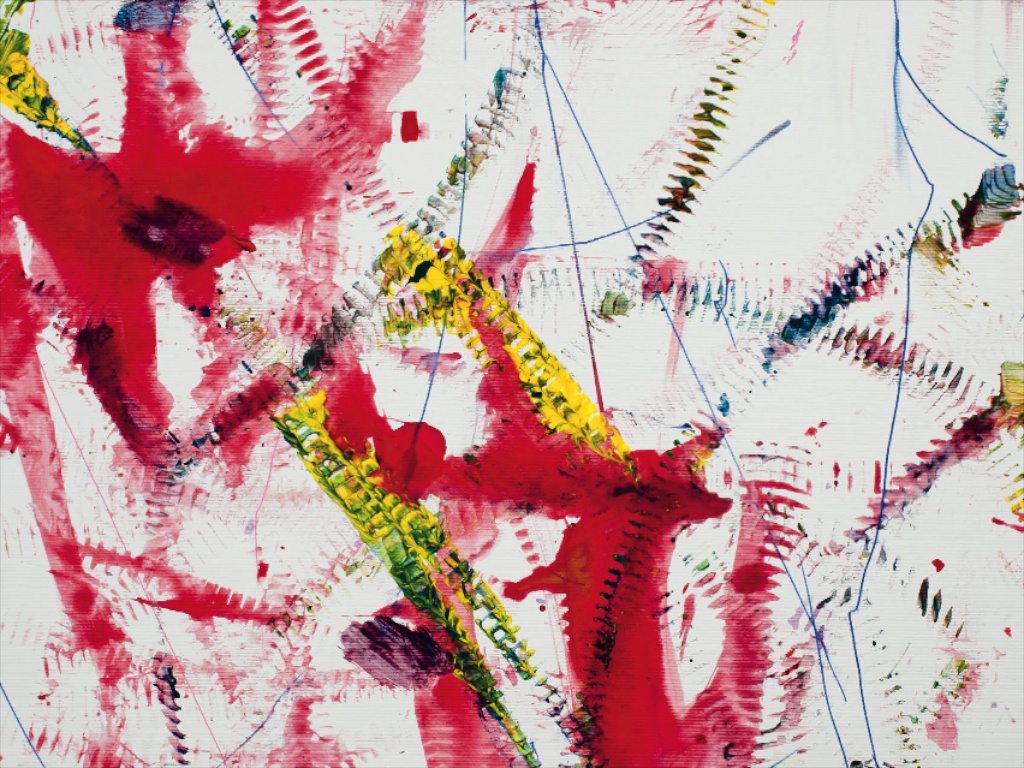}
    \end{subfigure}
    \caption{The exhibition with an ArtBots fight and some of the resulting paintings.}\label{fig:robot_paintings}
\end{figure*}

The last ArtBots day was dedicated to an exhibition, that was held at the Verbeke venue.
All 12 robots were presented by their makers to a broad audience of parents, teachers and sponsors.
The main event on the exhibition was a live painting session during which all painting robots were put on a huge canvas to co-create a painting.
In Figure~\ref{fig:robot_paintings} we show snapshots of the event, as well some of the resulting paintings.

\section{Results and discussion}\label{sec:discussion}
As stated in the introduction, the goal of our activity was threefold: (1) provide technical knowledge and robot building experience; (2) cause a cross-intercultural interaction and (3) train technology ambassadors who stimulate friends and other students to get involved in technology.
In order to evaluate whether these goals are met or not we observed our participants thoroughly during the activities and conducted a post-survey to question the participants about their experiences.
Our survey was completed by 26 out of 51 of the participants.

\subsection{Students perspective on difficulty}
\begin{figure*}[hbt]
    \centering
    \includegraphics[width=0.9\textwidth]{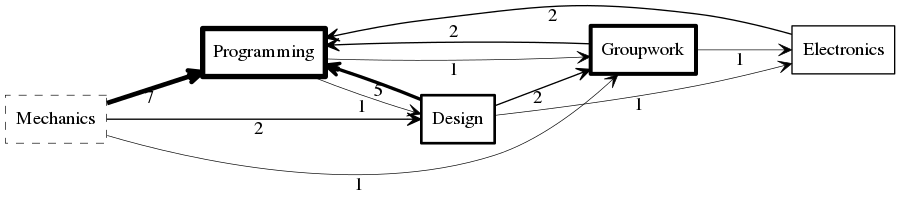}
    \caption{Difficulty of topics as perceived by the participants. The nodes show the topics that participants reported as easy or hard to learn. The arrows go from ``easy'' to ``hard'', with the width of the arrows indicating the percentage of participants that completed the survey (total number next to the arrow). Most students found programming difficult to learn and no student found mechanics difficult to learn.}\label{difficulty}
\end{figure*}

One of our goals was to provide technical knowledge and robot building experience.
While we did not measure the knowledge acquired by the students during our activity, we can say that all students participated in the building process and thus experienced the process of building robots.
In order to get acquainted with the students' perceived difficulties we asked them to tell us what they found hard and what easy.
The answers are summarized in Figure~\ref{difficulty} which shows that programming was perceived as the hardest to learn, while no student reported mechanics as hard to learn.
Electronics appears as a middle ground between these two.

Within the group of participants who completed the survey it seems that the more a skill departs from everyday experience, the hardest is to learn it.
This observation is anecdotal, as the amount of answers is not representative, and it coincides with the observation made by tutors and with the intuitive idea of learning being bootstrapped by the affordances\footnote{The affordances of an object are the opportunities it offers to interact with it, e.g. a piece of wood can be rotated by hand, texture and shape can be sensed by touch, it can be pushed/pulled/lifted, etc.; cf. the source code of a program.} provided by the objects manipulated~\citep{given2002}.
It might also be a consequence of the reduced time dedicated to practise this skill.

We read this perceived difficulty as an indication that the tools used in the learning of these skills need to be developed further to offer more affordances: to bring the skills and concepts closer to everyday experience.
This is also known as the virtual-to-real problem~\citep{dwengo2015_article}. In the mind of the participant, there are always expectations about what their robot should do (e.g. its behaviour).
These expectations can be expressed in natural language quite well, e.g. "When I press this button, my robot should start" or "When the light is too low the robot should turn around".
The participant can picture these ideas in their mind's robot simulator.
More often than not, these ideas are a simplification of the robot's real behaviour.
Consequently, when they need to bring these ideas down to the real hardware, some unforeseen aspects of the hardware will become important, making the translation from virtual to real less direct. 

We think that by integrating an intermediate learning step which involves adding abstraction layers to hardware and software, this virtual-to-real problem becomes easier to tackle for the participant.
For this we could for example include a robot simulator before switching to real robots. This kind of development are now common in physics, geometry and algebra~\footnote{\url{https://phet.colorado.edu/} and \url{https://www.geogebra.org/}}, but they are lacking in the field of robotics despite the huge advances brought by the open hardware community.

\subsection{Dealing with dangerous tools during robot construction}
Empowering teenagers to work with both hand and power tools requires a thoughtful approach.
During the live demonstration and try-out of the available tools, the students were asked to discuss the points of attention and safety rules.
Instead of imposing safety measures, it is important to let the students come up with some self-evident measures by themselves, such as wearing safety goggles and ear protection during cutting operations, only using paints, solvents or torches in well-ventilated rooms, etc.
Less comprehensible measures, such as not wearing safety gloves while using fast rotating power tools (e.g. miter saws, band sanders), were explained thoroughly.
An appropriate supervision of the teenagers is important to prevent injuries~\citep{Peterson1993}, alongside raising awareness of possible safety risks, especially in a mix of teenagers of strongly varying ages.

We draw a more general conclusion: giving responsibility to teenagers in handling dangerous situations, preceded by a discussion about safety in which there is room for the teenagers' own input, will have an overall positive effect on safety~\citep{Zaske2015}.
This corresponds with the observations we made during the activities.
All participants, including the youngest ones, handled the dangerous tools with care as instructed by the tutors.


\subsection{Observed social behaviour}

\begin{figure}[tb]
    \centering
    \includegraphics[width=0.45\textwidth]{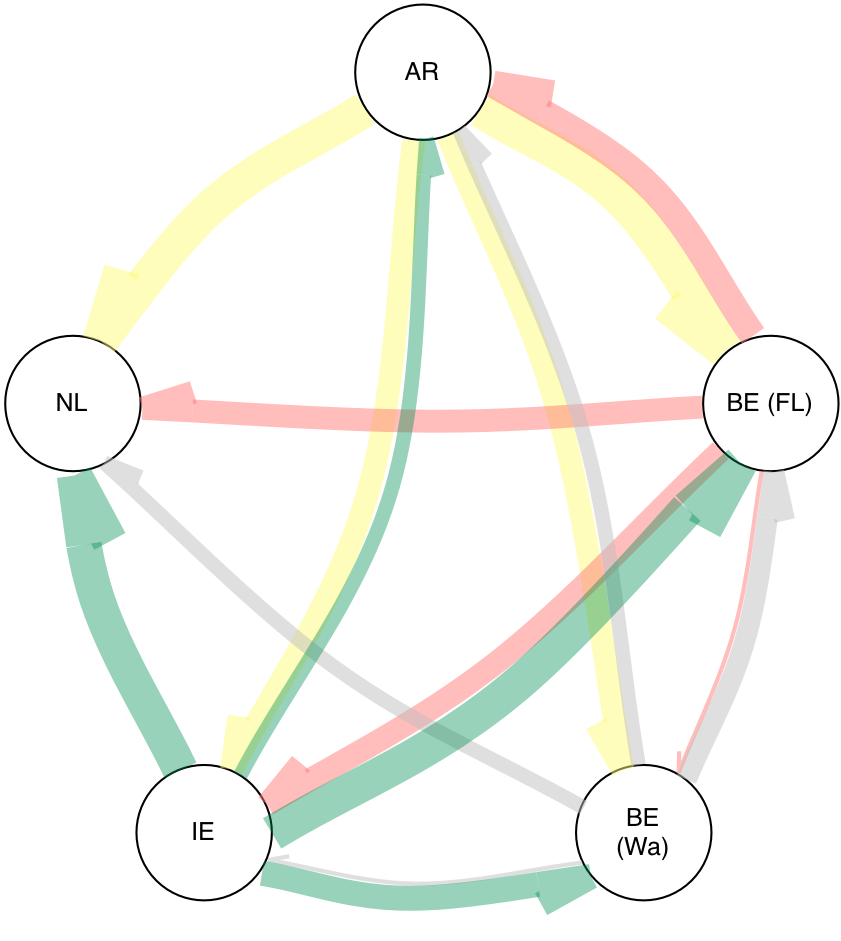}
    \caption{This infographic illustrates the interaction between participants from different countries. The thicker the arrow, the more students interacted with someone of a different nationality. Note that there are no leaving arrows from the Dutch students since they did not complete the surveys.}\label{fig:interaction}
\end{figure}

On the first day, participants grouped on small groups with people from their school.
However, after the brainstorm session the students were put together according the similarity of their ideas.
This resulted in groups of mixed nationalities and social backgrounds.
The social interaction as perceived by the participants (participants were asked with whom they had interacted) is visualised in Figure~\ref{fig:interaction}.
The thicker the arrow, the more students interacted with someone of a different nationality.
The size of the arrows were normalised with respect to the amount of students that completed our post-survey.
Note that there is no leaving arrow from the Dutch students, although they interacted with other students, simply because none of the Dutch students completed the survey.

By looking at the graph one notices immediately strong cultural differences in openness and willingness to interact with people from other cultures. 
For example, all Argentine students interacted with students from other nationalities.
This was despite their limited English proficiency and the fact that the Argentine students were the farthest from home.
On the other hand we noted limited interaction from the Walloon students which is confirmed by our survey.
Nevertheless we can say that our activity succeeded in achieving a cross-cultural collaboration between the participating students.
This is expressed greatly by the following quote of one of the Irish participants: {\it The most important lesson that I learned is that international barriers should not stop you from making friends. I learned this because I made new friends from Belgium, Netherlands and Argentina.}
The observed dynamics in the group brainstorming, or collaboratively generating ideas through idea sharing, supports the results of~\citet{wang2011} that introducing conceptual diversity, such as
cultural difference internal to a multicultural group, sparks more ideas and improves creativity.
\subsection{Becoming technology ambassadors}
An important aspect of the learning process is how enjoyable it is for the students.
It has been observed in the past~\citep{wyffels2010} that building robots can be used as a tool to learn how to solve problems which would be otherwise considered too abstract and tedious.
This was also confirmed during our activity.
The students were so engaged into building ArtBots that it was hard to convince them to go to a closing party with a DJ, drinks and candies (e.g. one of the social activities) on Sunday evening.
In our survey the participants graded the ArtBots event with $8.6$ points out of $10$.
Additionally, the students' answers in our post-survey reflect their enthusiasm about the activities:

\begin{itemize}
\item Un nuevo concepto de arte, que inspira para ver mas allá de los estándares propuestos; la creatividad puede llevarnos a construir robots originales y más útiles. {\it (Translation from Spanish) A new art concept that inspires to look beyond the standard; creativity can make us build orginal and more useful robots.}
\item I learned that with enthusiasm and compromise I can try to make a better world. The world can be more humane.
\item Als je met een groepje samenwerkt je, onder een beetje druk en tijdsgebrek, heel coole dingen kan maken! {\it (Translation from Dutch) When you work together in a team, under a little bit of time pressure, you can create very cool things!}
\item Il faut persister dans le travail que nous faisons et trouvé des solutions quand un problème se présente. {\it (Translation from French) By persisting in the work we found solutions whenever a problem occurred}
\item To open your eyes, there is more in the world than just internet.
\end{itemize}

\subsection{Final remarks}
The design and construction of a robot are not simple tasks and, unequivocally, the builder will face uncertainty. 
Under the right conditions this uncertainty provides a healthy amount of frustration and fuels curiosity.
The particular way in which the robot building task challenges students keeps them motivated and gives them impetus to overcome frustration.
This is the feature that makes robotic activities suited for many audiences and learning objectives.

A salient feature of robotic activities such as ArtBots, is that they induce inquisitive attitudes in the participants.
This was highlighted during the crash courses, see sections~\ref{program} and~\ref{electron}.
Admittedly, the task ``build a robot that makes art'' is vaguely defined.
Nevertheless, is this vagueness that gives power to the approach.
Granting that motivation is not an issue, when faced with an insurmountable lack of knowledge about something (maybe a tool, a sensor, the use of a device) students naturally engage in inquisitive behavior.
Students seek for the missing knowledge on the internet, ask their team mates or the tutors.
The benefit of this emergent behavior is twofold.
First, students train their abilities to pose question to others, question about extraordinary things, of which they might not know much about.
The process of formulating intelligible questions is fundamental in the process of understanding and inherent to the inquiry-based learning process~\citep{Colburn2000}.
This also endorses the effectiveness of interactive learning: you learn a lot more, faster and durable by doing it yourself than passively listening to a teacher~\citep{hake1998interactive,crouch2001peer}. 
Secondly, the tutors receive direct feedback on what topic cause the most problems for a group of students and/or what topic is most interesting for the group.
This helps with the steering of the ongoing event and provides extremely useful information for the design of future activities with similar student groups.
In terms of scientific literacy, the ArtBots project complies with the recommendations of~\citet{Kesidou2004} of showing relevance in science education and of paying attention to what students are thinking.

Another positive aspect of starting from an open problem or context, is that it removes structure from the learning process.
It is remarkable how the reduction of structure can spark creativity and inventiveness.
Although the reduction of structure can generate unwieldy frustration, it also allows each participant to adapt the learning process to their own pace, needs and curiosity.
The hands-on approach makes this possible: using tools as a means to an end, providing emergent learning experiences centered on the participants’ ideas and interests, creating the opportunity to explore one’s own creative potential.
Finally, by adopting the maker mindset, participants become producers instead of users, clients or consumers of technology.

Art is just one of the many non-technical disciplines that can be combined with engineering and science.
Fields related to history (e.g. archeology) could be combined in events when a technology of the past is being reproduced in the present; or when a technology that was crucial for a historical event (e.g. enigma in WWII) is being reproduced.
Similarly, literature can be combined with story telling and programming, advertisement and sociological models can be introduced using simulated agents or interactive machines~\citep{Bogost2010}.
The combinations are endless, we need only experiment and dare to break the pedagogical conventions.

\section{Conclusion}\label{sec:conclusion}
Taken as a whole, the underlying framework of this study suggests a fluid interaction among robot, student, and trainer.
Trainers facilitate access to robots and mediate related interactions with students, while robots are suitable for manipulations appropriate to the learning tasks.
Students engage in flexible, active, and integrated learning, they explore and test designs and solutions.
We conclude that by integrating arts and other social fields into STEM, one cannot only establish increased learning but also captivate a broader audience and tackle differences in age, gender and socio-cultural backgrounds.

However, one must be aware of the fact that such educational activities might ask of a higher commitment of educators.
A solution for this could be involving other actors such as parents with technical or artistic background.
Moreover, we also found that the acquisition of skills with fewer affordances (e.g. programming) require developed teaching tools, more time, and guidance. 

\section*{Acknowledgment}
The authors would like to thank all the volunteers and organisations\footnote{Please refer to \url{http://www.dwengo.org/projects/artbots} for a complete overview.} involved in the project. Without their support the ArtBots project would not have been possible. Additionally, we would like to thank all the participants for their enthusiasm and willingness to learn.
We would like to thank Cedric Verhelst for providing all the photographs in this paper.
The project was part of the Google RISE (Roots in Science and Engineering) 2014 program.

\section*{Author contribution}
All authors have contributed equally to the writing of this manuscript. The order of the authors list is inverse alphabetical; it is arbitrary and not a hierarchy.

\section*{Conflicts of interest}
Several authors are active members of the nonprofit organisations {\it Dwengo vzw} and {\it Dwengo Helvetica}.


\bibliographystyle{plainnat}
\bibliography{2016_artbots}

\end{document}